\documentclass[aps,twocolumn,floats,floatfix,amssymb,nofootinbib,prd,superscriptaddress]{revtex4-1}
\usepackage{graphicx,amssymb,amsmath,amsthm,amsfonts,epsfig}
\usepackage[linktocpage,colorlinks=true,citecolor=OliveGreen,linkcolor=Maroon,urlcolor=Maroon]{hyperref}
\usepackage{appendix}
\usepackage{tensor}
\usepackage[normalem]{ulem}
\usepackage[dvipsnames, usenames]{xcolor}
\usepackage{nicematrix}
\usepackage{xr-hyper}
\definecolor{linkcolor}{rgb}{0.0,0.3,0.5}
\usepackage[all]{hypcap}
\usepackage[T1]{fontenc}
\usepackage[utf8]{inputenc}
\usepackage[usenames,dvipsnames]{xcolor}

\definecolor{romared}{RGB}{142,0,28}

\newcommand{\be}{\begin{equation}}
\newcommand{\ee}{\end{equation}}

\def\be{\begin{equation}}
\def\ee{\end{equation}}
\newcommand{\beq}{\begin{eqnarray}}
\newcommand{\eeq}{\end{eqnarray}}

\usepackage{makecell}
\usepackage{soul}

\definecolor{cornellGreen}{HTML}{6EB43F}

\begin{document}
\title{Axion dissipation in conductive media and neutron star superradiance}
\author{Thomas F.M.~Spieksma}
\affiliation{Center of Gravity, Niels Bohr Institute, Blegdamsvej 17, 2100 Copenhagen, Denmark}
\author{Enrico Cannizzaro}
\affiliation{CENTRA, Departamento de F\'{\i}sica, Instituto Superior T\'ecnico -- IST, Universidade de Lisboa -- UL, Avenida Rovisco Pais 1, 1049 Lisboa, Portugal}
\begin{abstract}
In axion electrodynamics, magnetic fields enable axion-photon mixing. Recent proposals suggest that rotating, conductive plasmas in neutron star magnetospheres could trigger axion superradiant instabilities---an intriguing idea, given that such instabilities are typically associated with rotating black holes. In this work, we extend these investigations by properly incorporating plasma dynamics, particularly the plasma-induced photon effective mass, which suppresses the axion-photon mixing. Using two toy models for the conductive regions in the magnetosphere and accounting for fluid dynamics in both the frequency and time domain, we show that typical astrophysical plasma densities strongly inhibits the axionic instability. While our results assumes flat spacetime, the conclusions also apply to axion bound states in curved spacetimes, making neutron star superradiance less viable than previously thought. As a byproduct of our work, we provide a detailed description of axion electrodynamics in dissipative plasmas and uncover phenomena such as low-frequency axion ``tails'' and axion-induced electrostatic fields in dense plasmas.
\end{abstract}
\maketitle
\section{Introduction} 
The prediction of axions or axion-like particles is widespread in physics, as dark matter candidates~\cite{Bergstrom:2009ib,Marsh:2015xka,Hui:2016ltb,Ferreira:2020fam} or as a solution to the strong CP problem~\cite{Peccei:1977hh, Weinberg:1977ma,Wilczek:1977pj,Arvanitaki:2010sy}. One of the most promising ways to search for these particles is to exploit their coupling with the electromagnetic (EM) field. Due to this coupling, axions and photons can oscillate coherently in the presence of a sufficiently strong external magnetic field, in a similar fashion to neutrino oscillations~\cite{PhysRevLett.51.1415,PhysRevD.37.1237, Raffelt:1996wa, Mirizzi:2006zy}. The resulting phenomenology is extremely rich and has motivated many searches for axions, both on Earth, e.g., with light-shining-through-a-wall experiments~\cite{OSQAR:2015qdv, DellaValle:2015xxa, Ehret:2010mh} or axion haloscopes~\cite{PhysRevD.42.1297, PhysRevLett.104.041301, Ouellet:2018beu, PhysRevLett.118.091801}, and in astrophysical and cosmological contexts (see~\cite{Caputo:2024oqc, Safdi:2022xkm, OHare:2024nmr} for reviews on current constraints). Neutron stars (NSs) present a promising astrophysical playground as the extremely high magnetic fields facilitate axion-photon mixing. This potential has spurred a huge effort to utilize NSs as axion detectors~\cite{Huang:2018lxq, Hook:2018iia, Leroy:2019ghm,Battye:2019aco,Battye:2021xvt,Witte:2021arp, Millar:2021gzs,McDonald_2023,McDonald:2023shx,Safdi:2018oeu, Foster:2020pgt,Battye:2021yue,Foster:2022fxn,Battye:2023oac, Prabhu:2021zve,Noordhuis:2022ljw,Noordhuis:2023wid, Caputo:2023cpv,Long:2024qvd}. 
Recently, it was proposed that a magnetosphere-driven superradiant instability could occur for NSs as the magnetosphere contains a rotating, conductive plasma~\cite{Day:2019bbh}. Two key ingredients are needed for a superradiant instability:~a mode-confining mechanism and dissipative dynamics. The former is provided by the bare mass of the bosonic field, which allows bound state axion solutions around the star~\cite{Garbrecht:2018akc}, similar to the black hole case. However, whereas in black holes dissipation arises from the ergoregion, here it is provided by the conductivity in the rotating magnetosphere. The full process proposed in~\cite{Day:2019bbh} can be outlined as follows:~the NS magnetic field triggers photon production from the axion bound state, which then superradiantly scatters off the magnetosphere, extracting rotational energy from it. These amplified photons are then converted back into bound state axions, leading to an instability. The dissipative dynamics are thus contained in the plasma sector, which the axion can access indirectly through a ``portal'' provided by the\,photons.
In~\cite{Day:2019bbh} however, the dynamics of the magnetosphere were only partially taken into account. In particular, the plasma's linear response was modeled using Ohm's law with a real conductivity . In reality, the response of plasma to an EM perturbation consists of two components:~a conductive term that leads to mode absorption, and the harmonic motion of the plasma particles. The latter occurs at a frequency known as \emph{plasma frequency}, given by
\begin{equation}\label{eq:plasmafreq}
   \omega_{\rm p}= \sqrt{\frac{n_{\rm e} e^2}{m_{\rm e}}}\approx \frac{10^{-7}}{\hbar}\sqrt{\frac{n_{\rm e}}{10^{7}\text{cm}^{-3}}}\,\text{eV}\,,
\end{equation}
where $n_{\rm e}$, $m_{\rm e}$ and $e$ are the electron density, mass and charge, respectively, and we normalized to typical densities in NS magnetospheres~\cite{1969ApJ...157..869G}. As a result, the plasma can partially screen the propagating EM field, provided that the frequency of the photon is smaller than the plasma frequency, i.e., $\omega<\omega_{\rm p}$. In other words, the plasma endows the transverse polarizations of the photon with an effective mass given by the plasma frequency, such that their dispersion relation exhibits a gap $\omega^2=k^2+\omega_{\rm p}^2$, where $k$ is the wavenumber. 
\begin{table*}[t!]
    \centering
    \resizebox{0.98\textwidth}{!}{%
    \renewcommand{\arraystretch}{1.25}
    \begin{tabular}{|c|c|c|}
     \hline Scenario & Conclusion & Reference \\ \hline \hline
     $\sigma_{\rm par} = \nu = 0$ & Magnetosphere closed lines---Dense plasmas can suppress axion-photon mixing & Figure \ref{fig:nosigma}\,\&\,\ref{fig:EM_axion_plasma} \\ \hline 
     $\sigma_{\rm par} \neq 0\,,\nu = 0$ & Magnetosphere current sheets---Absorption of axion modes. Dense plasmas counter-balance this effect. & Figure \ref{fig:extconduct}  \\ \hline
     $ \sigma_{\rm par}=0\,,\nu \ll \mu$ & Accreting plasma---The axion decouples from the system when $\omega_{\rm p}\gg \mu$. Absorption is quenched. & Figure~\ref{fig:modeconduct}\,\&\,\ref{fig:EM_axion_fluid}  \\ \hline 
     $ \sigma_{\rm par}=0\,,\nu \gg \mu$ & Accreting plasma---The axion decouples on a scale set by the conductivity, i.e.,~$\sigma = \omega_{\rm p}^2/\nu \gg \mu$ & Figure \ref{fig:modeconduct}\,\&\,\ref{fig:Axion_highOmegap}  \\ \hline 
    \end{tabular}}
    \caption{Overview of the scenarios explored in this work and their outcomes. The parameter $\sigma_{\rm par}$ denotes the parametrized conductivity in Ohm's law, while $\nu$ represents the collision frequency. Throughout this work, conductivity is modeled either via collisions or Ohm's law---never both simultaneously.}
    \label{tab:conclusions}
\end{table*}
For non-conductive plasmas, it can be shown that in the limit $\omega_{\rm p} \gg \mu$ (where $\mu$ is the boson mass), axion-photon mixing is suppressed due to the disparity in the masses of the modes~\cite{PhysRevLett.51.1415,PhysRevD.37.1237, Mirizzi:2006zy}. Given that NS magnetospheres are filled with an extremely dense plasma, it is crucial to include this contribution in the system and determine whether the axion decouples from the photon---and thus from the plasma---thereby ``losing access'' to the dissipative dynamics.
The goal of this work is to assess the viability of axion superradiance in rotating NSs. Advances in NS modeling, combined with extensive pulsar observations, has enabled the use of NSs as axion detectors, yielding competitive constraints and opening an avenue to search for axions in the mass range [$10^{-8}$--$10^{-5}]\,\mathrm{eV}$~\cite{Noordhuis:2022ljw, Pshirkov:2007st, Battye:2019aco, Battye:2021yue, millar2021axionphotonUPDATED, Hook:2018iia, Foster:2020pgt, Gines:2024ekm, Tjemsland:2023vvc, Witte:2021arp, McDonald:2023shx, Mirizzi:2009nq, McDonald:2023ohd, Battye:2023oac, Battye:2021xvt, Caputo:2024oqc, Caputo:2023cpv}. In this context, superradiance offers a unique opportunity to probe \textit{lighter} axions than the range above using NSs. Since NSs are lighter than black holes, the relevant axion mass range for superradiance in NSs may differ from that associated with black holes~\cite{Cardoso:2018tly}. Additionally, the wealth of pulsar observations further enhances the prospects for this approach~\cite{Manchester:2004bp}.
This work is organized as follows. Section ~\ref{sec:realistic_env} outlines NS magnetospheres and superradiant clouds, including relevant scales and plasma dissipation. Section~\ref{sec:nonconductive_plasma} reviews axion-photon mixing without conductivity, while Section~\ref{sec:conductive_plasma} introduces the full plasma response, identifying regions where dissipative dynamics are important. We adopt two toy models: one with phenomenological conductivity via Ohm's law (Section~\ref{sec:parametric}) and another incorporating particle collisions (Section~\ref{sec:Drude}). Section~\ref{sec:overdense} examines \textit{over-dense plasmas}, where the plasma frequency exceeds the axion frequency, preventing on-shell photon production---a scenario relevant for axion bound states in plasma-filled NS magnetospheres. Our results show that whenever the plasma frequency is the dominant scale, axions decouple from the system and are unaffected by dissipative plasma dynamics. Finally, we discuss the implications of our results on NS superradiance in Section~\ref{sec:NSSR} and conclude in Section~\ref{sec:conclusions}. Our results are reported in terms of the boson mass $\mu$, we set $c = \hbar = 1$ and use rationalized Heaviside-Lorentz units for the Maxwell equations. Key conclusions are summarized in Table~\ref{tab:conclusions}.
\section{The system}\label{sec:realistic_env}
Given the complexity of the system, we first provide a general description of the mechanism and elucidate the relevant scales of interest.
\subsection{Axion bound states and superradiance}
Superradiant amplification occurs when bosonic modes scatter off a rotating, dissipative body, provided the following condition holds: 
\begin{equation}\label{eq:SRcondition}
\omega<m \Omega\,.
\end{equation}
Here, $\omega$ and $m$ are the frequency and the azimuthal number of the mode, respectively, and $\Omega$ is the angular velocity of the body. In our context, $\Omega$ corresponds to the rotational velocity of the magnetosphere, which co-rotates with the NS. 
Massive axions can form bound states around stars, potentially triggering a superradiant instability. Compared to black hole superradiance, $\Omega$ is typically much lower for NSs. Therefore, the fastest-spinning NSs, millisecond pulsars, have been identified as the most promising candidates for superradiance~\cite{Day:2019bbh,Kaplan:2019ako,Cardoso:2017kgn}.\footnote{Most millisecond pulsars are contained in binary systems, so that the cloud could also be disrupted by the tidal potential of the companion~\cite{Cardoso:2020hca,Baumann:2021fkf,Tomaselli:2023ysb,Tomaselli:2024bdd,Tomaselli:2024dbw}} For the fastest known millisecond pulsar, PSR J1748--2446ad~\cite{Hessels:2006ze}, which spins at $716\,\mathrm{Hz}$, the dominant $m=1$ mode satisfies the superradiant condition~\eqref{eq:SRcondition} with $\omega \approx \Omega \approx 0.088/r_{\rm s}$, where $r_{\rm s}= 2 M_{\rm s}$ is the Schwarzschild radius of the star, and $M_{\rm s} \simeq 2 M_{\odot}$ its mass. For axion bound states, the frequency of the superradiant mode is set by the axion mass $\omega \approx \mu$, yielding $\mu\approx 1.86 \times 10^{-11}\,\mathrm{eV}$. The resulting axion cloud peaks at a radius $\sim 1/r_{\rm s} \mu^2\approx 127 r_{\rm s}$. While higher or lower frequencies could be achieved by considering modes with larger $m$ or slower NS rotation, the corresponding instability timescales grow significantly, making the effect less relevant astrophysically.
\subsection{Pulsar magnetospheres}
The magnetosphere is filled with a dense, charge-separated plasma generated by particle extraction from the NS crust and pair production. This plasma screens the electric field produced by the star's rotation and carries currents along magnetic field lines. Since the plasma is frozen to these magnetic field lines, which are generated inside the star, it co-rotates with the NS up to a critical distance where the rotational velocity reaches the speed of light. This distance defines the \textit{light-cylinder}, $r_{\rm L}=1/\Omega$. Beyond this point, the plasma can no longer co-rotate with the star (see Figure~1 in~\cite{1969ApJ...157..869G}).
The region within $r<r_{\rm L}$, known as the ``near zone,'' is characterized by closed magnetic field lines. Beyond the light cylinder ($r>r_{\rm L}$) lies the ``wind zone,'' where the field lines open (i.e., closing only at large distances) and become more circular, allowing charged particles to stream outwards, forming the so-called ``wind''. In this region, the magnetic field is sustained by the particle currents rather than the star itself. 
Extensive theoretical studies and numerical simulations have shown that the bulk of the magnetosphere, where magnetic field lines are closed, is well described by the force-free condition, where EM fields combine to cancel the Lorentz force, resulting in a dissipationless plasma (see e.g. ~\cite{1969ApJ...157..869G, Komissarov:2005xc, Contopoulos:1999ga, Gralla:2014yja}). Dissipation becomes relevant only where the magnetic field opens up, forming current sheets where a non-vanishing resistivity is required to keep the model physically viable~\cite{Komissarov:2005xc,Palenzuela:2012my}.
The plasma density in the magnetosphere is generally proportional to the magnetic field strength and remains high overall. Typical electron densities range from $n_{\rm e}\!\sim\!\left[10^{6}-10^{12}\right]\,\mathrm{cm}^{-3}$(see Eqs.~(8)--(16) in~\cite{1969ApJ...157..869G}), corresponding to a plasma frequency in the range $\omega_{\rm p}\!\sim\!\left[10^{-7}-10^{-5}\right]\,\mathrm{eV}$.
Consequently, inside the magnetosphere, the condition $\omega_{\rm p}\gg \mu$ holds. To assess the interplay between the magnetosphere and the axion cloud, one must also compare their sizes. The magnetosphere’s outer boundary is set by the \emph{magnetospheric radius} $r_{\rm M}$ (also known as the ``Alfv\'{e}n'' radius). For an accreting NS---whether isolated and accreting from the interstellar medium or part of a binary---$r_{\rm M}$ can be estimated by equating the magnetic energy density with the kinetic density of the accreting material. Outside $r_{\rm M}$, the magnetic field is low enough such that accretion occurs gravitationally, thereby disrupting the magnetosphere.
The magnetosphere radius $r_{\rm M}$ depends on the NS parameters and the accretion rate. For strong magnetic fields, typically $r_{\rm M} \gtrsim r_{\rm L}$ holds, though in some cases, it may lie inside the light cylinder, eliminating the wind zone. In the pulsar regime instead, the Alfv\'{e}n radius can be large $r_{\rm M}> 12 r_{\rm s}$~\cite{2008AIPC.1068...87R, 2012MNRAS.420..216S}. Depending on the specific scenario, the axion cloud can extend far beyond the magnetosphere. Nevertheless, even in this case, the cloud remains immersed in the quasi-neutral plasma of the accretion flow. Estimating this plasma density is challenging due to its dependence on accretion dynamics, but it will always exceed that of the interstellar medium, where $\omega_{\rm p} \approx 10^{-10} \rm{eV}$~\eqref{eq:plasmafreq}. Thus, even outside the Alfv\'{e}n radius, one can safely assume $\omega_{\rm p}\gg \mu$.
\section{The theory} \label{sec:theory}
We now outline the equations governing our system. We consider a massive axion $a$ coupled to the EM field sourced by a cold plasma. The Lagrangian of this system reads
\begin{equation}\label{eq:lagrangian}
    \begin{aligned}
        \mathcal{L}&=-\frac{1}{2}\partial_{\alpha}a\partial^{\alpha}a
-\frac{1}{2}\mu^{2}a^{2}-\frac{1}{4}F_{\alpha\beta}F^{\alpha\beta}+A^\alpha j_\alpha\\&-\frac{g_{a \gamma \gamma}}{4}a\,{}^{*}\!F^{\alpha\beta}F_{\alpha\beta}\,,
\end{aligned}
\end{equation}
where $F_{\alpha\beta}=\partial_\alpha A_\beta-\partial_\beta A_\alpha$ is the Maxwell tensor, with $A_\alpha$ the EM potential and ${}^{*}\!F^{\alpha\beta}\equiv \frac{1}{2}\epsilon^{\alpha\beta\rho\sigma}F_{\rho\sigma}$ its dual. Furthermore, $\mu$ is the axion mass, $j_\alpha$ the EM plasma current and $g_{a \gamma \gamma}$ the axion-photon coupling constant. From the Lagrangian~\eqref{eq:lagrangian}, we obtain the equations of motion for the axion and Maxwell fields: 
\begin{equation}\label{eq:evoleqns}
\begin{aligned}
\left(\partial^\alpha \partial_\alpha-\mu^2\right) a &= \frac{g_{a \gamma \gamma}}{4}\,{ }^*\!F^{\alpha \beta} F_{\alpha \beta}\,,\\
\partial_\beta F^{\alpha \beta} &=j^{\alpha} - g_{a \gamma \gamma}{ }^*\!F^{\alpha \beta} \partial_\beta a\,.
\end{aligned}
\end{equation}
The modeling of the plasma and its conductivity enters the expression for the EM current in the right-hand side of Maxwell equations. The standard lore in many astrophysical scenarios is to assume a collisional plasma, a ``Drude model,'' i.e., a two-fluid electron-ion model including collisions between the two species. In this case, the conductivity arises naturally due to collisions between particles. However, this model is not suitable for NS magnetospheres, where the plasma is charge-separated and far from quasi-neutrality. Instead, the conductivity in current sheets is typically modeled using resistive magnetohydrodynamics, where the conductive term is described using Ohm's law~\cite{Komissarov:2005xc,Gruzinov:2007se,Palenzuela:2012my}. In the following, we adopt this approach, such that the current reads:
\begin{equation}\label{eq:Ohms_law_parametric}
j^\alpha=e (n_{\rm e} v_{\mathrm{e}}^\alpha-n_{\rm ion} v_{\mathrm{ion}}^\alpha)+\sigma_{\rm par}F^{\alpha \beta}v_{\mathrm{e},\beta}\,,
\end{equation}
where $n_{\rm ion}$ is the ion density, $v_{\mathrm{e/ion}}^\alpha$ is the electron/ion four velocity and $\sigma_{\rm par}$ is the parametrized conductivity. In Eq.~\eqref{eq:Ohms_law_parametric}, the first term captures the standard electron/ion response, while the second one is purely phenomenological and encapsulates the complex dynamics giving rise to a macroscopic conductivity in the NS magnetosphere~\cite{Li_2012,Kalapotharakos:2013sma,Brambilla:2015vta}.
To close the system~\eqref{eq:evoleqns}--\eqref{eq:Ohms_law_parametric}, we also consider the equations for the plasma velocity and density, i.e., the momentum and continuity equation. In the two-fluid formalism, electrons and ions are treated as two separate fluids. As electron-ion collisions also give rise to a macroscopic conductivity of the plasma, which may be relevant in the accreting quasi-neutral plasma outside the Alfv\'{e}n radius, we include them in the momentum equation. This can be done by considering a term proportional to the collision rate of the two species (see e.g.,~\cite{Dubovsky:2015cca}):
\begin{equation}\label{eq:continuity_momentum}
 v^\beta \partial_\beta v^\alpha=\frac{e}{m_{\rm e}}F^{\alpha\beta}v_{\mathrm{e},\,\beta}- \nu(v_{\rm e}^{\alpha}- v_{\rm ion}^\alpha), \quad \partial_\alpha (n_{\rm e} v^\alpha)=0\,,
\end{equation}
where $\nu$ is the collision frequency. Two analogous equations hold for the ion fluid. 
We will study axion-photon mixing in a conductive plasma in a uniform, constant magnetic field along the $\hat{y}$--direction, $B_y$. We consider the plasma to be static and isotropic, while we set the background axion field to zero. In addition, we will ignore ion perturbations throughout this work due to their large inertia compared to the electrons and consider them only as a neutralizing background. This setup is similar to a recent work~\cite{Cannizzaro:2024hdg}, which we refer to for details. Assuming plane waves to propagate in the $\hat{z}$-direction, the equations of motion~\eqref{eq:evoleqns},~\eqref{eq:Ohms_law_parametric} and \eqref{eq:continuity_momentum} reduce to a set of coupled, second order partial differential equations:
\begin{equation}\label{eq:simpleevolutionseqs}
\begin{aligned}
   \textbf{Axion}:&\quad \left(\partial_{t}^{2}-\partial_{z}^{2}+\mu^{2}\right)a -B_{y}g_{a\gamma\gamma}\partial_{t}A_{y}=0\,,\\
   \textbf{EM}:&\quad \left(\partial_{t}^{2}-\partial_{z}^{2}\right)A_{y}-e n_{\rm e} v_{y}-\sigma_{\rm{par}} \partial_t A_y \\&+ B_{y}g_{a\gamma\gamma}\partial_{t} a =0\,,\\
   \textbf{Fluid}:&\quad    
   \partial_{t}v_{y} -\nu v_{y}+ \frac{e}{m_{\rm e}}\partial_{t}A_{y}=0\,.
\end{aligned}
\end{equation}
We will typically show quantities rescaled by the electron charge-to-mass ratio e.g.~$\tilde{a} = (e/m_{\rm e})a$. More details are provided in Appendix~\ref{sec:numerical_procedure}.
\section{Non-conductive plasma}\label{sec:nonconductive_plasma}
To introduce axion-photon mixing in a simplified setup, we consider a non-conductive plasma, setting $\nu=\sigma_{\rm{par}}=0$. This scenario describes the bulk of the magnetosphere along closed field lines where plasma is dissipationless due to the force-free condition. The dynamics of the system are then fully captured by the so-called ``mass matrix,'' given by
\begin{equation}
\label{eq:Edispersion}  
\mathcal{M}=\frac{1}{2}
    \begin{bmatrix}
    -\omega_{\rm p}^2/ \omega &  B_y g_{a \gamma \gamma}  \\
    B_y g_{a \gamma \gamma} & -\mu^2/\omega 
    \end{bmatrix}\,.
\end{equation}
To simplify the problem, one can diagonalize this matrix through a rotation in the field basis by an angle
\begin{equation}\label{eq:mixing}
    \theta= \frac{1}{2}\,\text{arctan} \left(\frac{2 \omega B_y g_{a \gamma \gamma}}{-\omega_{\rm p}^2+\mu^2} \right)\,.
\end{equation}
This quantity is known as the \emph{mixing angle} and it characterizes the conversion probability between the axion and the photon as $P(a \leftrightarrow \gamma)\propto \sin^2 (2 \theta)$~\cite{Mirizzi:2006zy, PhysRevD.37.1237}. If either the magnetic field $B_y$ or the coupling $g_{a\gamma\gamma}$ vanishes, the mixing angle and consequently the conversion probability becomes zero. Crucially, a similar effect occurs when $\omega_{\rm p}\gg \mu$ and $\omega_{\rm p} \gg B_y g_{a \gamma \gamma}$. In other words, as the photon acquires an effective mass equal to the plasma frequency, the disparity of masses in the regime $\omega_{\rm p}\gg \mu$ heavily disfavours the conversion. Therefore, plasma oscillations induce an \emph{in-medium suppression} of the mixing~\cite{Cannizzaro:2024hdg}. Conversely, when $\mu=\omega_{\rm p}$, the conversion probability reaches its maximum at $\theta = \pi/4$, indicating a \emph{resonant conversion} between the two states. This is shown explicitly in Appendix~\ref{app:fullsystem_nocol}.

To gain more insight into the axion-photon system, we find the eigenfrequencies, which read
\begin{equation}\label{eq:omega1234}
\begin{aligned}
 \omega^{2}_{1,2}&= \frac{1}{2}\biggl(\omega_{a}^2+\omega_\gamma^2 +B_y^2 g_{a \gamma \gamma}^2\mp \\& \sqrt{\left(\omega_{\rm p}^2-\mu^2\right)^2+B_y^4 g_{a \gamma \gamma}^4+2 B_y^2 g_{a \gamma \gamma}^2 \left(\omega_{ a}^2+\omega_\gamma^2\right)}\biggr)\,, \\
\end{aligned}
\end{equation}
where we defined the free dispersion relations of the axion and the photon, respectively, as
\begin{equation}\label{eqn:freedispersion_relations}
    \omega_{a}^2\equiv k_z^2+\mu^2\,,\quad \omega_\gamma^2\equiv k_z^2+\omega_{\rm p}^2\,.
\end{equation}
Importantly, the eigenfrequencies~\eqref{eq:omega1234} do not correspond to the axion or photon modes individually, but rather to the frequency of a mixed state in a basis that diagonalizes the mass matrix~\eqref{eq:mixing}. Nevertheless, in the limit where the axion and photon decouple, we can identify them with the individual states.
\begin{figure}
    \centering
    \includegraphics[width = \linewidth]{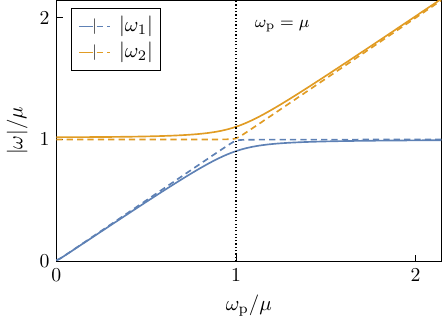}
    \caption{Eigenfrequencies of the axion-photon system in a magnetized, non-conductive plasma~\eqref{eq:omega1234}. Solid (dashed) lines indicate $B_y g_{a \gamma \gamma}/\mu=0.5$\,($0.05$). When $\omega_{\rm p}/\mu \ll 1$, $|\omega_{1}|$ ($|\omega_{2}|$) belongs to the ``pure'' axion (photon) state. For high plasma frequencies instead, the modes asymptote to the opposite state. The ``switch'' happens when $\omega_{\rm p} = \mu$, denoted by the black dotted line. We consider $k_z/\mu=0.01$ and $\mu=7$.}
    \label{fig:nosigma}
\end{figure}
This behavior is illustrated in Figure~\ref{fig:nosigma}, which shows the absolute value of the eigenfrequencies~\eqref{eq:omega1234} as a function of the plasma frequency. For $\omega_{\rm p}=0$ and moderate values of the coupling, the mixing between the axion and the photon is weak. In this regime, $|\omega_{1}|$ approximately corresponds to the axion mode with frequency $\omega \approx \mu$ and $|\omega_{2}|$ aligns with the photon mode with frequency $\omega\approx k_z$.\footnote{In the decoupling limit $g_{a \gamma \gamma} \rightarrow 0$, these modes reduce exactly to the free theory modes $|\omega_{1}|=\sqrt{k_z^2+\mu^2}$ and $|\omega_{2}|=k_z $.} As the plasma frequency increases and approaches $\omega \sim \mu$, the axion and photon modes oscillate, resulting in $|\omega_{1}|$ and $|\omega_{2}|$ becoming a combination of the photon and axion modes. This is evident in Figure~\ref{fig:nosigma}, where both eigenfrequencies grow parabolically with the plasma frequency in this intermediate regime. In the limit $\omega_{\rm p}\gg\mu$, the modes decouple again. Remarkably though, the eigenfrequencies asymptote to the opposite limit with respect to the vacuum case:~$|\omega_{2}|$ now corresponds to the axion state as it asymptotes to the free dispersion relation $|\omega_{2}|\rightarrow\sqrt{k_z^2+\mu^2}$, while $|\omega_{1}|$ grows parabolically with $\omega_{\rm p}$, as expected for photon modes. By decreasing the value of $g_{a\gamma\gamma} B_y$, the ``switch'' at $\omega_{\rm p}=\mu$ becomes sharper as modes oscillate less, while it becomes smoother for higher couplings. Finally, note that the eigenfrequencies are purely real as there is no conductivity providing dissipation.
\section{Conductive plasma}\label{sec:conductive_plasma}
We now consider conductive plasmas in two configurations. In Section~\ref{sec:parametric}, we model conductivity using Ohm's law to describe current sheets in the magnetosphere. In Section~\ref{sec:Drude}, we include collisional effects within a two-fluid plasma framework to capture the phenomenology outside the Alfv\'{e}n radius.
\subsection{Parameterized conductivity}\label{sec:parametric}
To isolate the impact of the plasma frequency and the conductivity on the superradiant instability, we treat both as free parameters, thereby keeping the plasma collisionless ($\nu = 0$). Analogous to the previous section, we determine the eigenfrequencies of the axion-photon system, now incorporating the parametrized conductivity:
\begin{equation}\label{eq:omega1234external}
\begin{aligned}
 &\omega^{2}_{1,2}=\frac{1}{2}\biggl(\omega_{ a}^2+\omega_\gamma^2 + B_y^2 g_{a \gamma \gamma}^2 -i\sigma_* \mp\\& \sqrt{\left(\omega_{\rm p}^2\!-\!\mu^2\!-\!i \sigma_*\right)^2\!+\!B_y^4 g_{a \gamma \gamma}^4 \!+\!2 B_y^2 g_{a \gamma \gamma}^2  \left(\omega_{a}^2\!+\!\omega_\gamma^2\!-\!i \sigma_*\right)}\biggr)\,, 
\end{aligned}
\end{equation}
where we defined $\sigma_*\equiv\sigma_{\rm par}\omega$. Due to the presence of conductivity, the eigenfrequencies of the system~\eqref{eq:omega1234external} are now complex. In the limit $\omega_{\rm p}\rightarrow \infty$, we can identify $|\omega_{1}|=\omega_{a}$, i.e., this mode reduces to that of the axion with a \emph{vanishing} imaginary part. This is a consequence of the axion and photon completely decoupling due to the photon's effective mass. Therefore, if the plasma frequency is the largest scale of the problem, as expected in a NS magnetosphere (see Section~\ref{sec:realistic_env}), the axion propagates freely, and is unaffected by the presence of a conductivity $\sigma_{\rm par}$. The plasma's response current $e n_{\rm e}v_{\mathrm{e}}^\alpha = \omega_{\rm p}^{2}m_{\rm e}v_{\mathrm{e}}^\alpha/e$ thus plays a crucial role in decoupling the axion. In the opposite limit, $\omega_{\rm p}, g_{a\gamma\gamma} \rightarrow 0$, $\omega_{1}$ reduces to the photon mode, similar to the previous section.
\begin{figure}
    \centering
    \includegraphics[width = \linewidth]{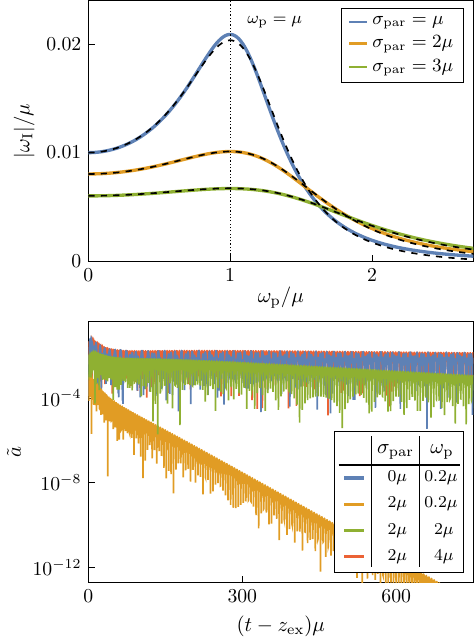}
    \caption{\textbf{Top Panel}:~The imaginary part of the ``axion mode'' (corresponding to $|\omega_{1}|$ (when $\omega_{\rm p}\gg \mu$) or $|\omega_{2}|$ (when $\omega_{\rm p}\ll \mu$) when considering axion-photon mixing in a plasma with an external conductivity $\sigma_{\rm par}$~\eqref{eq:omega1234external}. We pick $B_y g_{a\gamma\gamma}/\mu = 0.05$, $k_z/\mu = 0.01$ and $\mu = 7$. The black dashed lines indicate the estimate from Eq.~\eqref{eq:dielectric}. \textbf{Bottom Panel}:~The axion sector ($\tilde{a} = (e/m_{\rm e})a$) of the same system as above, but now in the time domain. We take $g_{a\gamma\gamma}B_y/\mu = 0.09$ and $\mu = 0.5$, while initializing at $z_0 = 5/\mu$ with $\mathrm{ID}_{a}$ and extracting at $z_{\rm ex} = 125/\mu$.}
   \label{fig:extconduct}
\end{figure}
The top panel of Figure~\ref{fig:extconduct} shows the imaginary part of the ``axion mode'' (described by $|\omega_{1}|$ and $|\omega_{2}|$ in the limits $\omega_{\rm p}\gg \mu$ and $\omega_{\rm p}\ll \mu$, respectively) for different values of the conductivity. In agreement with the previous observations, whenever $\omega_{\rm p} \gg \mu$, the imaginary part vanishes, independent of the external conductivity. The strength of the axion dissipation is thus completely regulated by the ratio of the axion mass to the plasma frequency $\mu/\omega_{\rm p}$:~as this ratio goes to zero, the imaginary part is increasingly suppressed. Near the resonant point $\mu \approx \omega_{\rm p}$, the imaginary part decreases for larger values of the conductivity $\sigma_{\rm par}$, reversing the hierarchy with respect to the $\omega_{\rm p} \gg \mu$ regime.
Interestingly, the imaginary part of the axion follows the estimate from~\cite{Lawson:2019brd,Caputo:2020quz, Berlin:2023ppd}, which studies axion absorption in dielectric media. In particular, using self-energy computations, the absorption rate $\Gamma$ is found to be:
\begin{equation}\label{eq:dielectric}
\Gamma \propto \omega_{\rm I} \simeq \frac{(B_{y}g_{a\gamma\gamma})^{2}}{\mu}\mathrm{Im}\left[\frac{-1}{\varepsilon(\omega)}\right]_{\omega=\mu}\,,
\end{equation}
where $\varepsilon(\omega)$ is the dielectric function. In an isotropic dielectric medium, it can be expressed as a function of the conductivity, $\varepsilon=1+i \sigma/ \omega$. To compare this with our scenario, we need to consider both the external conductivity and an imaginary conductivity that is proportional to the plasma frequency. This leads to $\varepsilon(\mu) = 1 - (\omega_{\rm p}/\mu)^2-i\sigma_{\rm par}\mu$, which reduces Eq.~\eqref{eq:dielectric} to
\begin{equation}
    \omega_{\rm I} \propto -\frac{(B_{y}g_{a\gamma\gamma})^{2} \mu^2 \sigma_{\rm par}}{(\mu^2-\omega_{\rm p}^2)^2+\mu^2\sigma_{\rm par}^2}\,.
\end{equation}
The above expression is indicated by black dashed lines in Figure~\ref{fig:extconduct} (\emph{top panel}) and shows excellent correspondence.
We evolve the same system in time domain~\eqref{eq:simpleevolutionseqs}. For details on our numerical framework, we refer the reader to Appendix~\ref{sec:numerical_procedure}. The system is initialized with a Gaussian wavepacket in the axion sector ($\mathrm{ID}_{a}$). The results, shown in Figure~\ref{fig:extconduct} (\emph{bottom panel}), confirm the intuition obtained in the frequency domain:~when conductivity is present, the imaginary part increases due to dissipation and the axion field (as well as the EM field) decreases (compare the blue and yellow line). However, as the electron density in the plasma increases (and thus the plasma frequency), the mixing between the axion and EM field is progressively suppressed, reducing the impact of the conductivity. In the regime $\omega_{\rm p} \gg \mu, \sigma_{\rm{par}}$, the axion and EM field effectively decouple, which can clearly be seen in Figure~\ref{fig:extconduct}, where the red and blue lines coincide.
\subsection{Toy model for translational superradiance}\label{sec:SR}
\begin{figure}
    \centering
    \includegraphics[width = 0.95\linewidth]{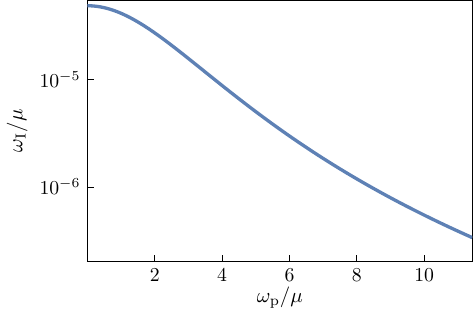}
    \caption{Imaginary part of the superradiant axion frequency as a function of the plasma frequency. We fix the parameters to $B_y g_{a\gamma\gamma}/\mu = 0.05$, $k_z/\mu = 3.571$, $\mu =7$, $v_{\mathrm{e}}^z=0.5$, and $c_{\rm s}=0.01$. As the plasma frequency increases, the instability is increasingly suppressed, illustrating the quenching effect of the plasma on the superradiance.}
    \label{fig:axion_SR}
\end{figure}
So far, we have examined the dissipation of the axion field in a static plasma. However, if the plasma possesses a non-zero velocity that exceeds the axion's phase velocity, the axion can instead be amplified through \emph{translational superradiance}, as demonstrated in~\cite{Day:2019bbh}. In this section, we explore the impact of the plasma frequency on this amplification. Following~\cite{Day:2019bbh}, we consider a drifting plasma with a background velocity component along the $\hat{z}$-axis, denoted by $v_{\mathrm{e}}^z$, and introduce a Lorentz-violating sound speed for the axion.\footnote{We note that a drifting plasma is not an exact solution to the background momentum equation when a homogeneous magnetic field is present. However, in the limit of small cyclotron frequency, this configuration serves as a good approximation, which we adopt here.} We repeat the analysis of the previous section and show the results in Fig.~\ref{fig:axion_SR}. As expected and in agreement with~\cite{Day:2019bbh}, we find superradiant modes whenever the superradiant condition~\eqref{eq:SRcondition} is satisfied. That said, our main conclusion remains unchanged:~increasing the plasma frequency drastically suppresses the instability timescale, rendering the effect negligible for any realistic astrophysical scenario.
\subsection{Collisional plasma (Drude model)}\label{sec:Drude}
We now consider the ``Drude model,'' where the conductivity is directly related to both the plasma and collision frequency, rather than being treated as a free parameter. Consider then photons propagating in a plasma in absence of axions ($g_{a\gamma\gamma}=0$) or background magnetic fields ($B_y=0$). From the momentum equation~\eqref{eq:continuity_momentum}, it can easily be seen that a static electron (and ion) fluid satisfies the background equations, even in the presence of collisions. Assuming again EM plane waves propagating along the $\hat{z}$-direction, we can solve the momentum equation in the frequency domain and obtain an expression for the perturbed velocities in the ($\hat{x},\hat{y}$)-direction:
\begin{equation}
    v_{x,y}=-\frac{e \omega}{m_{\rm e}}\frac{A_{x,y}}{i \nu + \omega}\,.
\end{equation}
Using this relation in the current $j_{x,y} = e n_{\rm e} v_{x,y}$, we find a complex conductivity described by Ohm's law:
\begin{equation}\label{eq:drude}
    j^{x,y}=\sigma E^{x,y}\,, \quad \text{with} \quad \sigma=\frac{\omega_{\rm p}^2}{\nu-i \omega}\,,
\end{equation}
where we introduced the electric field as $E^{x,y}=i \omega A^{x,y}$. As advertised, the conductivity is related to the plasma frequency and the collision frequency.
In the \emph{collisionless} limit ($\nu/\omega\rightarrow 0$), the motion of electrons is dominated by plasma oscillations, resulting in the conductivity~\eqref{eq:drude} becoming purely imaginary. This describes the effective mass of photons within the plasma. Conversely, when $\omega/\nu\rightarrow 0$, the motion of the electrons is controlled by collisions, leading to a purely real conductivity. In this regime, EM waves are damped due to Ohmic heating within the fluid. The \emph{Drude model} is thus able to capture both the effective mass of the photon and collisions, providing a consistent framework for testing the in-medium suppression by the plasma. For additional details on this model, we refer to Appendix~\ref{app:collisional}.
Even in presence of collisions, it is only the Maxwell equation in the $\hat{y}$--direction that couples to the axion. Similar to before, we obtain the characteristic eigenfrequencies as
\begin{equation}\label{eq:omega1234cond}
\begin{aligned}
 &\omega^{2}_{1,2}=\frac{1}{2}\biggl(\omega_{ a}^2+\omega_\gamma^2 + B_y^2 g_{a \gamma \gamma}^2 -\nu\sigma \mp\\& \sqrt{\left(\omega_{\rm p}^2\!-\!\mu^2\!+\!\nu\sigma\right)^2\!+\!B_y^4 g_{a \gamma \gamma}^4 \!+\!2 B_y^2 g_{a \gamma \gamma}^2  \left(\omega_{a}^2\!+\!\omega_\gamma^2\!-\!\nu \sigma\right)}\biggr)\,, 
\end{aligned}
\end{equation}
where we used Eq.~\eqref{eq:drude} for the standard conductivity of a Drude medium.\footnote{Strictly speaking, the medium is anisotropic due to the presence of a magnetic fieldm making the conductivity a tensor. In~\eqref{eq:omega1234cond}, we simply consider the component along the $\hat{y}$--direction, $\sigma_{yy}$.} The eigenfrequencies are complex due to the dissipation from the collisions. In fact, when $\nu\rightarrow 0$, they turn real and simply reduce to Eq.~\eqref{eq:omega1234}. Additionally, the modes decouple in the \emph{non-interacting limit} ($B_y g_{a\gamma\gamma} \rightarrow 0$) or the \emph{in-medium suppression limit} ($\omega_{\rm p}\rightarrow \infty$). For example, consider $|\omega_{1}|$:~(i) in the non-interacting limit, it corresponds to the dispersion relation of a transverse photon mode in a collisional plasma, representing a ``pure photon mode.'' (ii) Conversely, in the in-medium suppression limit, $|\omega_{1}|$ aligns with the axion free dispersion relation, representing a ``pure axion mode.'' The modes thus exhibit the same ``switch'' as in Figure~\ref{fig:nosigma}. Crucially, when the axion decouples from the system, it can no longer access the dissipative dynamics of the plasma. The scale at which this happens depends on specific properties of the plasma. We can distinguish two regimes:~the \emph{weakly-collisional} ($\mu \gg \nu$) or the \emph{strongly-collisional} ($\mu \ll \nu$) regime.
\begin{figure*}[t!]
    \centering
    \includegraphics[width = \linewidth]{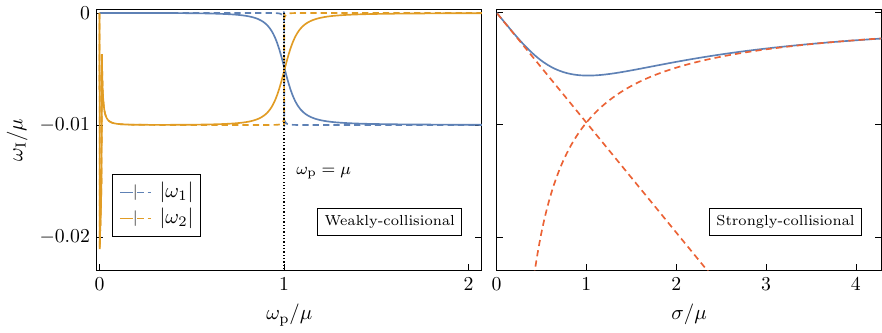}
    \caption{\textbf{Left Panel}: Imaginary part of the eigenmodes of the axion-photon system~\eqref{eq:omega1234cond} as a function of the plasma frequency. Solid (dashed) lines refer to $g_{a\gamma\gamma}B_y/\mu = 0.2\,(0.001)$, we take $\nu/\mu = 0.2$, $k_z/\mu = 0.01$ and $\mu = 7$. The peak on the left corresponds to a strongly-collisional plasma ($\nu\gg\omega$). \textbf{Right Panel}: The imaginary part of the axion mode when considering axion-photon mixing in a collisional plasma. We ensure to be in the collision dominated regime by taking $\nu/\mu = 7.5$ and $\mu = 7$. Moreover, we take $k_z/\mu = 0.01$, and $B_y g_{a\gamma\gamma}/\mu = 0.035$. The red dashed lines follows the prediction from Eq.~\eqref{eq:im_part_strongly_col}.}
    \label{fig:modeconduct}
\end{figure*}
\subsubsection{Weakly-collisional plasma}\label{sec:weakly-collisional}
In the weakly-collisional regime, EM modes are sourced by the axion with $\omega \gtrsim \mu \gg \nu$, resulting in oscillatory motion of the electrons within the plasma. As can be seen in Eq.~\eqref{eq:drude}, the conductivity in this regime is mostly imaginary. In the left panel of Figure~\ref{fig:modeconduct} we show the imaginary part of the eigenmodes~\eqref{eq:omega1234cond}.\footnote{Note that the real part of the eigenmodes follows exactly the same trend as in the collisionless case, and can thus be found in Figure~\ref{fig:nosigma}.} As before, the modes decouple when $\omega_{\rm p}\gg\mu$, with $|\omega_{2}|$ taking the role of the axion. Crucially, upon decoupling at $\omega_{\rm p} \geq \mu$, the imaginary part of the axion mode drops completely. The smaller the coupling $g_{a\gamma\gamma} B_y$, the sharper this drop (see dashed lines). The scale that regulates the presence of an imaginary part (and therefore of an instability) is thus the ratio between the plasma frequency and the axion mass, rather than between the conductivity and the axion mass. Finally, the imaginary part of the axion mode in the in-medium suppression regime ($\omega_{\rm p}\gg \mu$) can be well described by
\begin{equation}\label{eq:im_part_weakly_col}
    \omega_{\rm I}\approx- \nu \frac{ g_{a\gamma\gamma}^2 B_y^2}{2\omega_{\rm p}^2}\,.
\end{equation}
%
\subsubsection{Strongly-collisional plasma}
In the strongly-collisional regime ($\nu \gg \mu \approx \omega$), the collision frequency dominates over the plasma oscillations, and the plasma effectively behaves as a conductor. The conductivity~\eqref{eq:drude} is now mostly real, and given by $\sigma\approx \omega_{\rm p}^2/\nu$. In this regime, and assuming to be in the ``weak-mixing limit'' ($g_{a\gamma\gamma}B_y \ll \mu$), the imaginary part of the axion modes is found as~\cite{Ahonen:1995ky}\footnote{The solution in~\cite{Ahonen:1995ky} at $\sigma\ll\mu$ is missing a factor $2$. We correct this error in Eq.~\eqref{eq:im_part_strongly_col}.}
\begin{equation}\label{eq:im_part_strongly_col}
    \omega_{\rm I} \approx -\frac{g_{a\gamma\gamma}^2 B_y^2}{2} 
    \begin{cases}
    \begin{aligned}
      &\sigma/\mu^2\!&&\text{for}\ \,  \sigma \ll \mu\,,\\
      &1/\sigma\!&&\text{for}\ \, \sigma \gg \mu\,,
    \end{aligned}
    \end{cases}
\end{equation}
which are denoted in Figure~\ref{fig:modeconduct} (\emph{right panel}) by the red dashed lines. Note that for $\sigma \gg \mu$, the lower expression in Eq.~\eqref{eq:im_part_strongly_col} coincides with the one in the weakly-collisional regime~\eqref{eq:im_part_weakly_col}. It shows that the dynamics in both regimes are in essence the same, with the key difference being the parameter that sets the scale:~the plasma frequency $\omega_{\rm p}$ or the conductivity $\sigma = \omega^2_{\rm p}/\nu$. Consequently, in the strongly-collisional regime, the  imaginary part of the axion is regulated by the ratio between the conductivity $\sigma=\omega_{\rm p}^2/\nu$ and the axion mass. Remarkably, in the limit where $\sigma \rightarrow \infty$, the imaginary part of the axion decreases rather than increases. This may seem counter-intuitive, yet the underlying physical mechanism was explored in~\cite{Ahonen:1995ky}, drawing an analogy to the ``Zeno paradox'' in quantum mechanics. Specifically, the dissipation caused by the conductivity can be understood on the quantum level as the plasma absorbing photons, where each absorbed photon acts as a ``measurement'' of the system. When these ``measurements'' occur frequently (as is the case in the strongly-collisional regime), they prevent the evolution of the initial axion state, as each one collapses the wavefunction (and thus reset the system’s state). As a result, the conversion rate from axions to photons is inversely related to the ``measurement rate,'' i.e., the conductivity~\eqref{eq:im_part_strongly_col}.
\subsubsection{Time domain analysis}
\begin{figure}
    \centering
    \includegraphics[width = 0.95\linewidth]{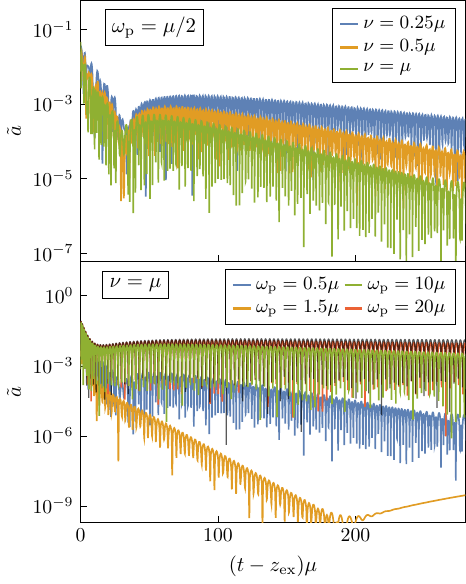}
    \caption{Axion sector when evolving the axion-photon system in a (weakly-)collisional plasma. \textbf{Top panel}:~We show the impact of the conductivity for $\omega_{\rm p}/\mu = 0.5$. \textbf{Bottom panel}:~We show the impact of the plasma frequency for $\nu = \mu$. In addition, we show a ``free axion'' with $g_{a\gamma\gamma} = 0.0$, denoted by the black line. In both panels, $g_{a\gamma\gamma}B_{y}/\mu = 0.225$, $\mu = 0.2$ and we use $\mathrm{ID}_{a}$, which we start at $z_0 = 2/\mu$, while we extract at $z_{\rm ex} = 50/\mu$. The late-time behavior of the yellow curve is related to an induced drift of the electrons in the plasma and further detailed in Appendix~\ref{app:absorb_photons}.}
    \label{fig:EM_axion_fluid}
\end{figure}

The plasma frequency dictates the presence of an axionic instability in collisional plasmas; specifically, when the plasma frequency exceeds all the other scales, the imaginary part of the axion is \emph{always} suppressed. Depending on the configuration, this suppression arises at two different scales
\begin{equation}
    \begin{cases}
    \begin{aligned}
      &\mu \ll \omega_{\rm p}\!&&\text{when}\ \,  \nu \ll \mu\,,\\
      &\mu \ll \omega_{\rm p}^2/\nu\!&&\text{when}\ \, \nu \gg \mu\,.
    \end{aligned}
    \end{cases}
\end{equation}
To solidify our conclusions, we perform a time domain analysis of this system. We consider the system of equations~\eqref{eq:simpleevolutionseqs}, setting $\sigma_{\rm par}=0$. As the Drude model entails a frequency-dependent conductivity, one cannot simply resort to Ohm's law in the time domain and we must evolve the fluid velocity. In what follows, we will again initialize a Gaussian wavepacket in the axionic sector $\mathrm{ID}_{\rm a}$.
Consider the regime $\nu \ll \mu$. As shown in Figure~\ref{fig:EM_axion_fluid} ({\it top panel}), the presence of collisions leads to a \emph{dissipation} of axion modes, which are exponentially suppressed at low plasma frequencies. However, as shown in Figure~\ref{fig:EM_axion_fluid} ({\it bottom panel}), increasing $\omega_{\rm p}$ interrupts this trend. In the limit $\omega_{\rm p}\gg \mu$, the axion decouples from the system, completely quenching its dissipation, consistent with the frequency domain analysis. In fact, for high enough plasma frequencies (red line), the evolution coincides with that of a free axion (black line) ($g_{a\gamma\gamma} = 0$). Notice that the axion solutions in Figure~\ref{fig:EM_axion_fluid} also present a low frequency ``tail'' component emerging after the absorption of the high-frequency modes. This is visible in the yellow line in the bottom panel and, although not shown, holds for all other curves at later times. Interestingly, these tails are generated by stationary drift solutions of EM modes in collisional plasmas, which we further detail in Appendix~\ref{app:absorb_photons}.
Finally, in Appendix~\ref{sec:Stronglycollisional} we consider also the strongly collisional regime $\nu\gg\mu$ in the time domain. In agreement with the frequency-domain analysis, Figure~\ref{fig:Axion_highOmegap} shows how in this case the presence of dissipation is regulated by $\sigma=\omega_{\rm p}^2/\nu$. Furthermore, we provide details on the collision-induced phenomenology in axion-photon mixing in Appendix~\ref{app:collisional}.
\section{Dynamics in an over-dense plasma}\label{sec:overdense}
\begin{figure}
    \centering
    \includegraphics[width = 0.95\linewidth]{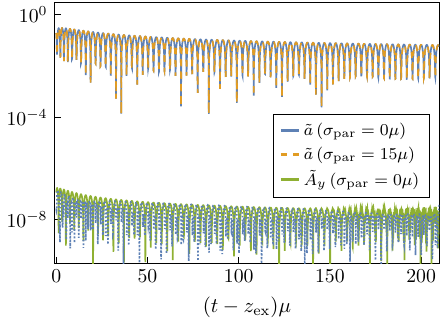}
    \caption{We show the axion and EM sector for $\omega_{\rm p} = 30\mu$, $g_{a\gamma \gamma}B_y/\mu = 0.0003$ and $\mu = 0.3$. We initialize at $z_0 = 4.5/\mu$ and extract at $z_{\rm ex} = 30/\mu$. While the axion field is unaffected by the presence of a conductivity (blue vs.~yellow line), the EM field is completely suppressed in presence of a conductivity. Additionally, by rescaling the axion signal (blue dotted line) according to Eq.~\eqref{eq:electrostatic}, we obtain a perfect agreement with the electro-static field (green line).}
    \label{fig:electrostatic}
\end{figure}

Thus far, we have explored {\it dynamical mixing}, where the frequency of the modes is the dominant scale in the system, i.e., $\omega > \omega_{\rm p}, \mu, \sigma$. This naturally raises the question:~what happens when an on-shell axion with $\omega>\mu$ propagates through an over-dense plasma with $\omega_{\rm p}>\omega$? Since the mixing occurs in a linear theory, any photon generated by the axion must share its frequency ($\omega \sim \mu$). This suggests that, in such a regime, the axion cannot produce on-shell photons. Nonetheless, as we demonstrate below, while  no on-shell photons are created in this regime, the axion can still induce a non-propagating electrostatic field in the plasma. However, the formation of this field is also suppressed due to in-medium effects.
Consider Eqs.~\eqref{eq:simpleevolutionseqs} in the regime $\omega<\omega_{\rm p}$, assuming $\sigma_{\rm{par}}=\nu=0$ for simplicity. In the frequency domain, the EM field admits a solution with $\partial_z A_y=0$, given by
\begin{equation}
\label{eq:electrostatic}
    A_y=\frac{i B_y g_{a\gamma\gamma}\,\omega\,a}{\omega^2-\omega_{\rm p}^2}\,.
\end{equation}
This is an axion-induced \textit{electro-static} field.\footnote{This field resembles the electro-static solution found in~\cite{Gines:2024ekm} along the longitudinal direction.} It can only be generated within a magnetic field and it cannot propagate. Figure~\ref{fig:electrostatic} shows this electro-static field as it arises in our time-domain simulations. To confirm whether the electric field is indeed static, we consider a simulation where the background magnetic field is shut off at large radii. Although not shown explicitly in Figure~\ref{fig:electrostatic}, the results show that an electric field is generated in the $B_y\neq 0$ region, but does not propagate into the $B_y=0$ zone, confirming the electro-static nature of the signal. Additionally, turning on conductivity (yellow line) shows no effect on the axion field. Finally, we compare the induced electro-static field (green line) with the prediction from Eq.~\eqref{eq:electrostatic} (dotted blue), finding excellent agreement. 
\section{On Equations in Curved Spacetime }\label{sec:curved}
In curved spacetime, the axion couples at leading order to the axial degree of freedom of the photons. On Schwarzschild, the master equation for the axial photon mode in a conductive plasma reads~\cite{Day:2019bbh, Cardoso:2017kgn}:
\begin{equation}
\label{eq:curved}
    \left[-\frac{\mathrm{d}^2}{\mathrm{d}r_*^2}-\omega^2+\frac{\ell(\ell+1)}{r^2}-i \sigma (\omega-m \Omega)\right]r A_{\ell m}=0\,,
\end{equation}
where $\ell$ and $m$ are the angular indices, $r_*$ is the tortoise coordinate and $A_{\ell m}$ denotes the radial part of the axial wavefunction. Reference~\cite{Day:2019bbh} suggested that the imaginary part of the conductivity leads to a tachyonic instability in the superradiant regime as the superradiant factor $\omega-m \Omega$ flips sign. In what follows, we argue that this interpretation does not hold. Instead, similar to the flat spacetime case discussed above, the imaginary part of the conductivity $\mathrm{Im}(\sigma)$ gives rise to an effective mass term for the photon.
The key point is that $\mathrm{Im}(\sigma)$ is frequency-dependent, unlike $\mathrm{Re}(\sigma)$. For a \emph{static} plasma, the dielectric function takes the form $\epsilon = 1 - \omega_{\rm p}^2/\omega^2$, such that the imaginary conductivity is given by $\mathrm{Im}(\sigma) = \omega_{\rm p}^2/\omega$. However, in a \emph{drifting} or \emph{rotating} plasma, this expression is modified. For example, for a plasma drifting with velocity $v$, one must replace $\omega \rightarrow \omega - \mathbf{k} \cdot \mathbf{v}$ in these expressions. This follows from a coordinate transformation, as detailed in Sec.~4.3 of~\cite{KrallTrivelpiece1973}, which shows that in the rest frame of the electrons, the disturbance is at the plasma frequency. A similar argument applies in the case of a rotating plasma:~the relevant frequency transforms as $\omega \rightarrow \omega - m \Omega$. Hence, Eq.~\eqref{eq:curved} can be rewritten as:
\begin{equation}
\begin{aligned}
    \Bigg[&-\frac{\mathrm{d}^2}{\mathrm{d}r_*^2}-\omega^2+\frac{\ell(\ell+1)}{r^2}\\& -i \mathrm{Re}(\sigma)(\omega-m \Omega)+\omega_{\rm p}^2\Bigg]r A_{\ell m}=0\,,
\end{aligned}
\end{equation}
indicating that the photon acquires an effective mass $\omega_{\rm p}$---just as in the flat spacetime case---and the same phenomenology holds. It is worth noting that this shift does not apply for the real part of the conductivity, as it is frequency-independent and remains invariant under boosts or rotations.
A similar conclusion can be obtained from a thermal field theory calculation. The equation of motion for a photon field can be written as
\begin{equation}
    \partial^2 A^\mu(x)+\int \mathrm{d}^4 x\, \Pi_{\rm R}^{\mu\nu}(x, x') A(x')=0\,,
\end{equation}
where $\Pi^{\mu\nu}$ denotes the retarded photon self-energy. Following~\cite{Chadha-Day:2022inf}, this can be expanded as
\begin{equation}
    \partial^2 A^\nu+ \mathrm{Re}[\Pi_{\rm R}^{\mu \nu}(q,x)]_{q=0}A_\mu-\partial_q^\rho \mathrm{Im}[\Pi_{\rm R}^{\mu \nu}(q,x)]_{q=0}\partial_\rho A_\mu=0\,,
\end{equation}
where $\Pi_{\rm R}$ is the Wigner transform of the self-energy. Neglecting spatial dispersion in the medium’s response function and using the relation between the self-energy and conductivity, this expression reduces in the temporal gauge ($A^0=0$) to
\begin{equation}\label{eq:thermal}
    \partial^a A^i+\omega_{\rm{p}}^2 A^i+\text{Re}[\sigma(0)]\partial_t A^i=0\, ,
\end{equation}
where we used $\text{Im}[\omega \sigma(\omega)]_{\omega\rightarrow 0}=\omega_{\rm p}^2$. For a Drude model (as in Eq.~\eqref{eq:drude}), $\text{Re}[\sigma(0)]=\omega_{\rm{p}}^2/\nu$, allowing the real part of the conductivity to be identified with the damping rate, and the imaginary part with the effective mass. From Eq.~\eqref{eq:thermal}, it is evident that in a rotating frame, where $\partial_t\rightarrow \partial_t-\Omega \partial_\varphi$, rotation affects the damping term but leaves the mass term unchanged. As a result, no tachyonic instability arises when $\omega-m \Omega$ changes sign. 
\section{Neutron Star Superradiance}\label{sec:NSSR}
We now discuss the implications of our results for NS superradiance. As outlined in Section~\ref{sec:realistic_env}, the plasma in both the magnetosphere and the accretion flow satisfies the condition $\omega_{\rm p} \gg \mu$. In Sections~\ref{sec:nonconductive_plasma} and~\ref{sec:conductive_plasma}, we demonstrated that photon production is entirely suppressed under this condition, causing the axion field to effectively ``lose access'' to the dissipative plasma dynamics.
Although the axion cloud cannot produce on-shell photons, an axion-induced electric field could still form near the rotating magnetosphere (see Section~\ref{sec:overdense}). This field might, in principle, extract rotational energy and transfer it to the axion. However, the high plasma density within the magnetosphere suppresses the formation of such a field (see Eq.~\eqref{eq:electrostatic}), effectively preventing superradiance.\footnote{While photons in overdense plasmas can form bound states around compact objects~\cite{Cannizzaro:2020uap, Cannizzaro:2021zbp}, these states are extremely fragile and geometry-dependent~\cite{Dima:2020rzg}, so we do not consider them further.}
In summary, the plasma within the magnetosphere is expected to be dense enough to effectively screen the electric field and suppressing the superradiance mechanism. Although the plasma density decreases outside the magnetosphere, it remains sufficiently high due to accretion. Morerover, the magnetic field weakens outside the Alfv\'{e}n radius ($B \propto r^{-1}$~\cite{1969ApJ...157..869G}), diminishing the prospects for superradiance in this region as well.
Finally, we estimate the conductivity of the magnetosphere by matching pulsar models~\cite{2012ApJ...746...60L, 2012ApJ...749....2K} with observations. For example, Refs.~\cite{Kalapotharakos:2013sma, Brambilla:2015vta} compared dissipative magnetosphere pulsars models with data obtained from Fermi-LAT, finding good agreement for values in the range $0.01 \Omega \lesssim \sigma \lesssim 100\Omega$. Assuming $\Omega \sim \omega \sim \mu$ from the superradiant condition~\eqref{eq:SRcondition}, this corresponds to $\sigma \approx 10^{-11}\,\mathrm{eV}$ for millisecond pulsars. Given that $\omega_{\rm p} \gg \sigma$, we conclude that plasma-induced suppression severely quenches the axion growth rate in pulsars.
The suppression scale is set by the condition $\mu<\omega_{\rm p}$, i.e.,
\begin{equation}
     \mu < 10^{-7} \sqrt{\frac{n_{\rm e}}{10^{7}\,\mathrm{cm}^{-3}}}\,\mathrm{eV}\,,
\end{equation}
which clearly includes the relevant superradiant mass range. 
In order to have a viable superradiant instability, the superradiance growth rate must be faster than the pulsar spin-down timescale~\cite{Cardoso:2017kgn}. Even in the absence of plasma effects, the superradiant rate was already found to be higher than the spin-down rate~\cite{Day:2019bbh}. Our results show that plasma dynamics introduce an additional suppression factor, rendering axion superradiance highly unlikely under realistic astrophysical conditions.
\section{Conclusions}\label{sec:conclusions}
Superradiance is a widely studied phenomenon, from both fundamental and observational perspectives (see~\cite{Brito:2015oca} and references therein). Its occurrence depends on the existence of a dissipative mechanism, which for black holes is naturally provided by the horizon. When considering superradiance in other astrophysical scenarios, alternative dissipation mechanisms must be present~\cite{Richartz:2013unq,Cardoso:2015zqa,Cardoso:2017kgn,Day:2019bbh}. For NSs, it has been proposed that conductive plasmas in the magnetosphere could play this role and facilitate axion superradiance~\cite{Day:2019bbh}. In this work, we further explore this possibility by accounting for the plasma dynamics.
To maintain generality, we model the plasma conductivity using two distinct approaches:~(i) a parameterized, purely phenomenological model and (ii) the Drude model, where conductivity arises from collisions within the plasma. In both approaches, we find that the axionic instability is suppressed when the plasma frequency dominates all other relevant scales in the system. Moreover, by considering typical parameters for NS magnetospheres, we demonstrate that the axionic instability---and thus the superradiance mechanism---is always suppressed in realistic scenarios. 
Although our results are derived in flat spacetime, we expect that the conclusions will also hold in curved spacetime. The primary difference would be a change in the sign of the imaginary part of the axion, which would then turn unstable{, as in the toy model of Sec.~\ref{sec:SR}}. However, the suppression we find is related to the magnitude of the imaginary part, not its sign. As a result, we expect that, even in curved spacetime, the plasma frequency increases the timescale of the instability, making it irrelevant from an observational point of view. 
Finally, beyond superradiance, our findings offer insights into the ability of axions to heat up baryonic matter. This may have important consequences in early-universe cosmology and galactic dynamics, similar to the discussions in~\cite{Ahonen:1995ky,Dubovsky:2015cca}. 

\vspace{0.5cm}
\noindent {\bf Acknowledgments.} 
We thank Andrea Caputo for initially suggesting to study the impact of realistic plasmas on neutron star superradiance and for collaborating at various stages of this work. We are grateful to Sam Witte for the helpful comments during the preparation of the manuscript and to Vitor Cardoso and Yifan Chen for useful discussions. We also thank the anonymous
referee for useful suggestions aimed at improving our manuscript. The Center of Gravity is a Center of Excellence funded by the Danish National Research Foundation under grant No. 184. We acknowledge support by VILLUM Foundation (grant no. VIL37766), the DNRF Chair program (grant no. DNRF162) by the Danish National Research Foundation, and the European Union’s H2020 ERC Advanced Grant ``Black holes: gravitational engines of discovery'' grant agreement no. Gravitas–101052587.
\appendix
\section{Numerical procedure}\label{sec:numerical_procedure}
In this appendix, we outline the specifics of the time-domain simulations. The equations of motion for the full system are given in Eq.~\eqref{eq:simpleevolutionseqs}. To favour a stable numerical evolution, we apply two modifications to these equations, similar to a previous work~\cite{Spieksma:2023vwl}:~(i) Plasma responds to an EM perturbation proportional to the electron charge-to-mass ratio, which is extremely large. To avoid numerical issues, we rescale the fields by this ratio, e.g.~$\tilde{a} = (e/m_{\rm e})a$. Since we are dealing with linear perturbations, such amplitudes can be rescaled freely. (ii) We reduce the second-order wave equations to first order equations, making use of the ``conjugate momentum'' $P_{x} \equiv (\partial_{t}+\partial_{z})x$. The resulting evolution equations~\eqref{eq:simpleevolutionseqs} then read
\begin{equation}\label{eq:EOM}
\begin{aligned}
   \textbf{Axion:}\quad \partial_{t}P_{\tilde{a}}&=\partial_{z}P_{\tilde{a}}-\mu^{2}\tilde{a} +B_{y}g_{a\gamma\gamma}\partial_{t}\tilde{A}_{y}\,,\\
   \partial_{t}\tilde{a} &= P_{\tilde{a}} - \partial_{z}\tilde{a}\,, \\
   \textbf{EM:}\quad \partial_{t}P_{\tilde{A}_{y}}&=\partial_{z}P_{\tilde{A}_{y}}+\omega_{\rm p}^{2} v_{4}- B_{y}g_{a\gamma\gamma}\partial_{t}\tilde{a}\,,\\
   \partial_{t}\tilde{A}_{y} &= P_{\tilde{A}_{y}} - \partial_{z}\tilde{A}_{y}\,,\\
   \textbf{Fluid:}\quad    
   \partial_{t}v_{4}&= \nu v_{4}- \partial_{t}\tilde{A}_{y}\,,\\
\end{aligned}
\end{equation}
where we have made use of the definition of the plasma frequency~\eqref{eq:plasmafreq}.
We consider two types of initial data, initializing either the axion ($\mathrm{ID}_{a}$) or EM sector ($\mathrm{ID}_{\rm EM}$) with a Gaussian wavepacket. In particular, we take initial conditions $a(0,r) \equiv a_0$ and $A_{y} (0,r) \equiv A_{y0}$ with
\begin{equation}\label{eq:ID}
\begin{aligned}
\mkern-22mu \begin{pmatrix}
a_0\\
A_{y0}
\end{pmatrix} &= \begin{pmatrix}
\mathcal{A}_{a}\\
\mathcal{A}_{\rm EM}
\end{pmatrix}\exp{\left[-\frac{(z-z_0)^2}{2 \sigma^2}-i\Omega_{0} (z-z_0)\right]} \,, \\[8pt]
\partial_t a_0 &= -i\Omega_0 a_0\,, \quad \partial_t A_{y0} = -i\Omega_0 A_{y0}\,,
\end{aligned}
\end{equation}
where $(\mathcal{A}_{a}, \mathcal{A}_{\rm EM})^\top=(1,0), (0,1)$ for $\mathrm{ID}_{a}$ and $\mathrm{ID}_{\rm EM}$, respectively. For the width of the wavepacket, we choose $\sigma = 3.0/\mu$, yet our results do not depend on this factor. For the frequency, we typically choose $\Omega_0 = 0.4\mu$, which should be larger than the plasma frequency when initializing with $\mathrm{ID}_{\rm EM}$ in order for the modes to propagate.
We evolve~\eqref{eq:EOM} in time with initial conditions~\eqref{eq:ID} using a two-step Lax-Wendroff scheme \cite{Krivan_1997,Pazos_valos_2005,Zenginoglu:2011zz,Zenginoglu:2012us,Cardoso:2021vjq,Cannizzaro:2024yee,Cannizzaro:2024hdg} where we employ second-order finite differences. Our grid size is uniformly spaced in $z$ with $\mathrm{d}z = 0.06$, and $\mathrm{d}t = 0.5\,\mathrm{d}z$ such that the Courant–Friedrichs–Lewy condition is satisfied at all times. Convergence tests of the numerical routine are reported in an earlier work~\cite{Cannizzaro:2024yee}.
\section{Collisional plasmas:~details and consistency checks}\label{app:collisional}
In this appendix, we provide additional details on the Drude model and the results of Section~\ref{sec:Drude}. First, in Section~\ref{app:absorb_photons}, we turn off the couplings between the axion and photon and study the impact of the plasma on the EM sector. Then, in Section~\ref{app:fullsystem_nocol}, we evolve the full axion-EM-fluid system, however we turn off the collisions in order to test the impact of the plasma on the axion-photon conversion. Finally, in Section~\ref{sec:Stronglycollisional}, we provide additional details on the strongly-collisional regime.
\subsection{Absorption of photons}\label{app:absorb_photons}
\begin{figure}[htbp]
    \centering
    \includegraphics[width = 0.95\linewidth]{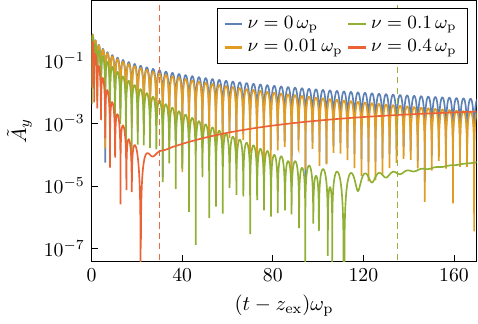}
    \caption{EM field when evolving the EM-fluid system, without couplings to the axion ($g_{a\gamma\gamma}B_{y}/\omega_{\rm p} = 0$). We set $\omega_{\rm p}  = 0.1$ and initialize a purely EM Gaussian wavepacket ($\mathrm{ID}_{\rm EM}$) at $z_0 = 1/\omega_{\rm p}$, while we extract at $z_{\rm ex} = 25/\omega_{\rm p}$. As the collision frequency is increased, the EM field loses more energy to the plasma. The vertical dotted lines show where the zero-frequency component approximately kicks in.}
    \label{fig:EM_fluid_conductivity}
\end{figure}

We analyze the EM-fluid system~\eqref{eq:EOM} in the absence of couplings to the axion sector, i.e., $g_{a\gamma\gamma} = 0$, in order to isolate the effects of conductivity on the system. As illustrated in Figure~\ref{fig:EM_fluid_conductivity}, increasing the collision frequency leads to a suppression of the amplitude of the EM field. In other words, when the collision frequency is higher, photons lose more energy to the plasma, heating it up. We can compute the heat generation $Q$ due to this ``Ohmic heating'' through
\begin{equation}
    Q = \mathbf{J} \cdot \mathbf{E} = \sigma |\mathbf{E}|^{2}\,,
\end{equation}
where $\mathbf{J}$ is the current density and $\mathbf{E}$ the electric field. The time-averaged heat dissipation rate is then
\begin{equation}
    \langle Q \rangle = \frac{1}{2} \mathrm{Re}\left(\sigma\right) |E_0|^{2} = \frac{1}{2} \frac{\omega_{\rm p}^{2}\nu}{\nu^{2} + \omega^{2}} |E_0|^{2}\,,
\end{equation}
where we used Eq.~\eqref{eq:drude} for the conductivity and $E_0$ is the strength of the electric field, which in our case is the amplitude of the initial Gaussian. Note that the heat generation is maximized when $\nu \sim \omega$.
In addition, there is a zero-frequency component emerging at late times, similar to Figure~\ref{fig:EM_axion_fluid}. As we increase the collision frequency, this component starts to dominate at progressively earlier times. To understand why, we should have a closer look at the Drude model. Under some external EM field $\mathbf{E}$, the response from the electrons in the plasma goes as
\begin{equation}\label{eq:response_Drude}
    m_{\rm e}\frac{\mathrm{d}\mathbf{v}_{\rm e}(t)}{\mathrm{d}t} = -e\mathbf{E}-m_{\rm e}\nu \mathbf{v}_{\rm e}(t)\,.
\end{equation}
Before solving this equation, we note that besides the induced oscillations, there exists a steady-state (zero-frequency) solution ($\mathrm{d}\mathbf{v}_{\rm e}/\mathrm{d}t = 0$), the so-called ``drift velocity'' given by
\begin{equation}\label{eq:drift}
    \mathbf{v}_{\rm drift} = -\frac{e\mathbf{E}}{m_{\rm e}\nu}\,,
\end{equation}
which induces a current that goes like
\begin{equation}
    \mathbf{J} = - n_{\rm e}e \mathbf{v}_{\rm drift} = -\frac{\omega_{\rm p}^{2}}{\nu}\mathbf{E}\,.
\end{equation}
While the initial response from the plasma is to damp the high frequency modes of the EM perturbation (exponentially, like $e^{-\nu t}$), there thus exist another solution, independent of time. The higher the collision frequency, the earlier the oscillatory modes are damped and the other solution can dominate. 
To make this more quantitative, we can find the full response~\eqref{eq:response_Drude} as
\begin{equation}
\mathbf{v}(t)=v_0e ^{-\nu t}\cos{\omega t}+\frac{eE_0}{m_{\rm e}\nu}(1-e^{-\nu t})\,.
\end{equation}
The zero-frequency part should thus start to dominate when
\begin{equation}
\tau_{\rm dom} = \frac{1}{\nu}\mathrm{ln}\left(\frac{v_{0}m_{\rm e}\nu}{e E_0}\right)\,.
\end{equation}
The ratio between timescales for two simulations, say $\nu_1$ and $\nu_{2}$, can thus be found as 
\begin{equation}
\frac{\tau_{\mathrm{dom}, \nu_1}}{\tau_{\mathrm{dom}, \nu_2}} \approx \frac{\nu_2}{\nu_1}\left(1+\frac{\mathrm{ln}(\nu_1)-\mathrm{ln}(\nu_2)}{\mathrm{ln}(x)}\right)\,,
\end{equation}
where $x \equiv v_{0}m_{\rm e}/(e E_0)$ and we have assumed $\mathrm{ln}(\nu_{1}) \ll \mathrm{ln}(x)$. From the red and green curve Figure~\ref{fig:EM_fluid_conductivity}, we immediately see that the above estimate at leading order holds:~the zero-frequency component starts to dominate about $4$ times later for the green curve compared to the red one, indicated by the vertical dotted lines. We have verified this scaling extensively for different values of $\nu$ and longer timescales, and it remains consistent. 
\subsection{Without collisions}\label{app:fullsystem_nocol}
\begin{figure}[h!]
    \centering
    \includegraphics[width = 0.95\linewidth]{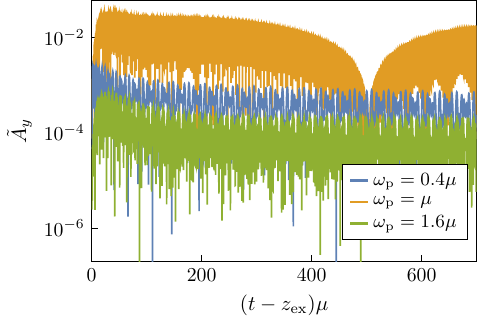}
    \caption{EM sector when evolving the full system, without any conductivity ($\nu/\mu = 0$). We set $g_{a\gamma\gamma}B_{y}/\mu = 0.025$ and initialize a purely axion Gaussian wavepacket ($\mathrm{ID}_{\rm a}$) at $z_0 = 5/\mu$, while we extract at $z_{\rm ex} = 125/\mu$. We take the axion mass as $\mu = 0.5$. The conversion from axions to photons is suppressed for higher plasma frequencies, except at the resonant frequency $\omega_{\rm p} = \mu$ where production is maximized.}
    \label{fig:EM_axion_plasma}
\end{figure}

We now consider the full axion-EM-fluid system~\eqref{eq:EOM}, excluding collisions ($\nu = 0$) to focus on the role of the plasma frequency in axion-photon conversions. As shown in Figure~\ref{fig:EM_axion_plasma} and discussed in Section~\ref{sec:conductive_plasma}, higher plasma frequencies lead to a suppression of the photon production. In addition, as anticipated, when the axion mass matches the plasma frequency ($\mu = \omega_{\rm p}$), a resonant conversion occurs, maximizing photon production.
\subsection{Strongly-collisional plasma}
\label{sec:Stronglycollisional}
\begin{figure}[h!]
    \centering
    \includegraphics[width = 0.95\linewidth]{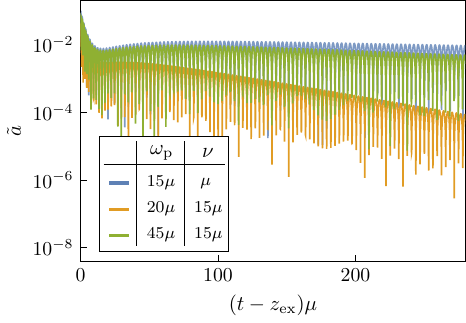}
    \caption{Axion sector when evolving the full system. We show $\mu = 0.2$ and $g_{a\gamma \gamma}B_y/\mu = 0.225$, we use $\mathrm{ID}_{\rm a}$, which we start at $z_0 = 2/\mu$, while we extract at $z_{\rm ex} = 50/\mu$.}
    \label{fig:Axion_highOmegap}
\end{figure}

Whereas in Figure~\ref{fig:EM_axion_fluid}, we focused on the weakly-collisional regime, here we consider the full Drude model in the strongly-collisional regime. In Figure~\ref{fig:Axion_highOmegap}, we show a time domain evolution. The results are again consistent with the observation from the frequency domain:~when increasing the collision frequency the axion amplitude decreases due to energy being lost to the fluid. Yet again, by increasing the plasma frequency, the axion loses access to the dissipative dynamics and the conversion of axions to photons is disfavoured.
\vspace{0.5\textheight}
\bibliography{ref}
\end{document}